# Broadband Sub-λ GHz ITO Plasmonic Mach-Zehnder Modulator on Silicon Photonics

Rubab Amin,[1] Rishi Maiti,[1] Yaliang Gui,[1] Can Suer,[1] Mario Miscuglio,[1] Elham Heidari,[2] Ray T. Chen,[2] Hamed Dalir,[3] and Volker J. Sorger[1,*]

[1]Department of Electrical and Computer Engineering, George Washington University, Washington, DC 20052, USA

[2]Microelectronics Research Center, Electrical and Computer Engineering Department, University of Texas at Austin, Austin, Texas 78758, USA

[3]Omega Optics, Inc. 8500 Shoal Creek Blvd., Bldg. 4, Suite 200, Austin, Texas 78757, USA

*Corresponding author: sorger@gwu.edu

**Here, we demonstrate a spectrally broadband, GHz-fast Mach-Zehnder interferometeric modulator, exhibiting a miniscule $V_\pi L$ of 95 V·μm, deploying a sub-wavelength short electrostatically tunable plasmonic phase-shifter, based on heterogeneously integrated ITO thin films into silicon photonics.**

Indium tin oxide (ITO), belonging to the class of transparent conductive oxides, is a material extensively adopted in high-tech industry such as in touchscreen displays of smartphones or contacts for solar cells. Recently, ITO has been explored for electro-optic modulation using its free-carrier dispersive effect enabling unity-strong index modulation [1-4]. However, GHz-fast modulation capability using ITO is yet to be demonstrated – a feature we show herein. Given the ubiquitous usage of phase-shifter technologies, such as in data communication, optical phased arrays, analog and RF photonics, sensing etc.; here we focus on a Mach-Zehnder interferometer (MZI) based modulator to demonstrate a comprehensive platform of heterogeneous integration of ITO-based opto-electronics into silicon photonic integrated circuits (PIC). Since for phase-shifters only the real-part of the optical refractive index ($n$) is of interest, in previous studies we have shown the interplay between a selected optical mode (e.g. photonic bulk vs. plasmonic) and the material's figure of merit ($\Delta n/\Delta \alpha$), where $\alpha$ is the optical loss, directly resultant form Kramers-Kronig relations [5]. Additionally, ITO can be selectively prepared (via process conditions [6]) for operating in either an $n$-dominant or $\alpha$-dominated region [5]. Using this approach, we recently showed a photonic-mode ITO-oxide-Si MZI on silicon photonics characterized by a $V_\pi L$ = 0.52 V·mm [3], and a plasmonic version deploying a lateral gate exhibiting a $V_\pi L$ = 0.063 V·mm [7]. Indeed, a plasmonic mode enables a strong light-matter-interaction (e.g. extrinsic slow-light effect), which, when superimposed with ITO's intrinsic slow-light effect, proximal epsilon-near-zero (ENZ) effects [8], enables realization of just 1-5 μm short phase-shifters [5]. The device-advantage of such micrometer-compact opto-electronics are small capacitances, in the order of ~fF, enabling low power consumption and small RC delays for fast signal reconfiguration, as shown here. Note, ENZ operation incurs high losses detrimental for phase shifters; as such, the desired operation region is adequately close-to, but not at the high-loss ENZ (ENZ located in $\alpha$-dominant region) [5]. Here, we discuss an ITO-plasmon-based phase-shifter hetero-geneously integrated into a silicon photonic MZI delivering GHz-fast modulation while being spectrally broadband, thus opening up opportunities for multi-spectral operation. Refraining to exploit feedback from a resonator has several advantages: a) no spectral alignment to the pump laser is required; b) Thus, no heaters are needed for thermal tuning, which would raise the power consumption per-device into the ~nJ/bit levels [9]; c) Heat spreads ~100 μm's across the PIC, thus packaging density can be significantly improved; d) The photon-lifetime of high-$Q$ cavities can limit RF modulation speed, which is not the case here.

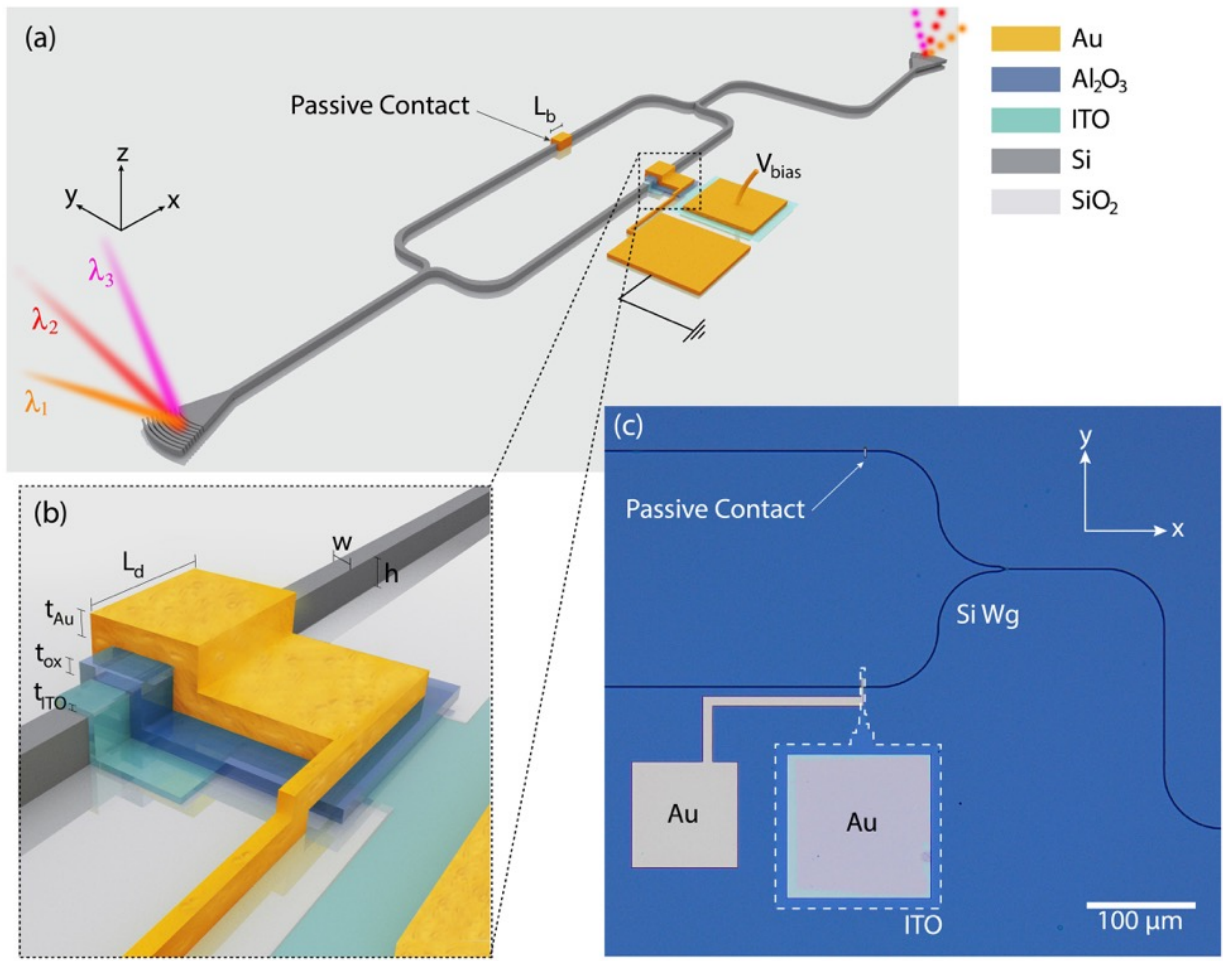

**Figure 1.** (a) Schematic of the broadband GHz plasmonic ITO-based Mach Zehnder modulator ; (b) Active device region, $L_d$; $t_{Au}$ = 50 nm, $t_{ox}$ = 20 nm, $t_{ITO}$ = 10 nm, w = 500 nm, h = 220 nm; corresponding FEM eigenmode profiles to light ON and OFF states (inset); (c) Optical microscope image of the sub-λ ($L_d$ = 1.4 μm) modulator.

The base interferometer is taped out as a symmetric SOI MZI to minimize chirp effects in modulation induced by different splitting ratios in the Y-junctions of the MZI and includes post-tape out loss-balancing between both arms using a metallic strip ($L_b$) on the non-modulated arm to minimize extinction ratio (ER) degradation effects (Fig. 1a). Our sweep of the active phase-shifter device length ($L_d$) ranges from sub-$\lambda$ (1.4 µm) to $\lambda$-scale devices (3.5 µm) [Fig. 1b]. Broadband spectral response is measured in C-band region (~30 nm, Fig. 2a); which is expected since the plasmonic resonance of the mode has a FWHM of 100's of nm. The spectral response of the device is determined by ITO dispersion and proximity to ENZ. For ultra–broadband applications (e.g. 100+ nm) ITO modulators for different spectral regions (e.g. $\Delta\lambda$ = 50 nm) can be processed using different conditions as demonstrated earlier [6]. Functional capacitor traits in the measured bias range is observed (Fig. 2b). DC electro-optic transmission power tests and squared cosine fit (as dictated by MZI operating principle) result in an ER of ~3 dB to >8 dB, respectively (Fig. 2c).

The measured $V_\pi L$ is just 95±2 V·µm and found to be rather constant across all device scaling. While these prototype devices require a relatively high voltage ($V_{\pi,\text{as-is}}$ ~26 V, $\epsilon_{r,Al_2O_3}$ = 11, $t_{ox}$ = 20 nm), the $V_{\pi,\text{future-option}}$ = 1 V can be reached using 5 nm high-k dielectrics (e.g. $Hf_3N_4$ [10]) in a push-pull configuration. The results indicate a modal index change $\Delta n_{eff}$ of ~0.2 (Fig. 2d) and FEM eigenmode analysis (inset, Fig. 1a) reveals an ITO index change of about 0.6 (Fig. 2e) reflecting an ~2× increased modal confinement factor ($\Gamma$) corresponding to active biasing, slightly lower than previous modulators in ITO [3], and intentionally enabling lower insertion loss (IL) of about 6.7 dB. Cutback measurements reveal 1.6 dB/µm propagation loss in the active capacitive stack and an additional 1.3 dB/coupling loss from in/out coupling of the mode from the Si waveguide, while the passive contact for loss balancing (Fig. 1a, $L_b$) exhibits a 1.2 dB/µm propagation loss and 1.1 dB/coupling loss correspondingly. The deposited ITO thin film carrier concentration, $N_c$ of 3.1×10$^{20}$ cm$^{-3}$ is determined from metrology (e.g. ellipsometry) and a carrier change $\Delta N_c$ = 2.1×10$^{20}$ cm$^{-3}$ estimated from the gated measurements places the operation of these devices in the *n*-dominant operation region, however intentionally away from the high-loss ENZ (6-7×10$^{20}$ cm$^{-3}$) state, yet sufficiently near to capture a slow-light effect [5].

Frequency response ($S_{21}$) is obtained by generating a low power modulating signal (0 dBm) with a 50 GHz network analyzer; a bias-tee combines DC voltage bias (6 V) with the RF signal (Fig. 2f). RF output from the modulator is amplified using broadband EDFA (~35 dB), an optical tunable filter enhances the signal integrity and reduces undesired noise by 20 dB. The modulated light is collected by a photodetector with a single mode fiber. The -3 dB roll-off (small signal) shows a speed of 1.1 GHz (Fig. 2g), which matches estimations for the RC-delay given a capacitance of 213 fF (area = 42 µm$^2$) and a total resistance of 680 $\Omega$, aligned with our earlier results [3]. A proportionate dynamic switching energy of ~21 pJ/bit characterizes the spectral broadband tradeoff in this device. Future improvements such as optimized contact placement, deployment of high-k gate dielectrics including $t_{ox}$ scaling, and utilizing push-pull schemes can enable switching energy reduction of ~10$^3$×, thus only requiring a few fJ/bit. However, aJ/bit energy levels are likely not feasible in non-resonator schemes due to the tradeoff in spectral bandwidth (i.e. no cavity feedback).

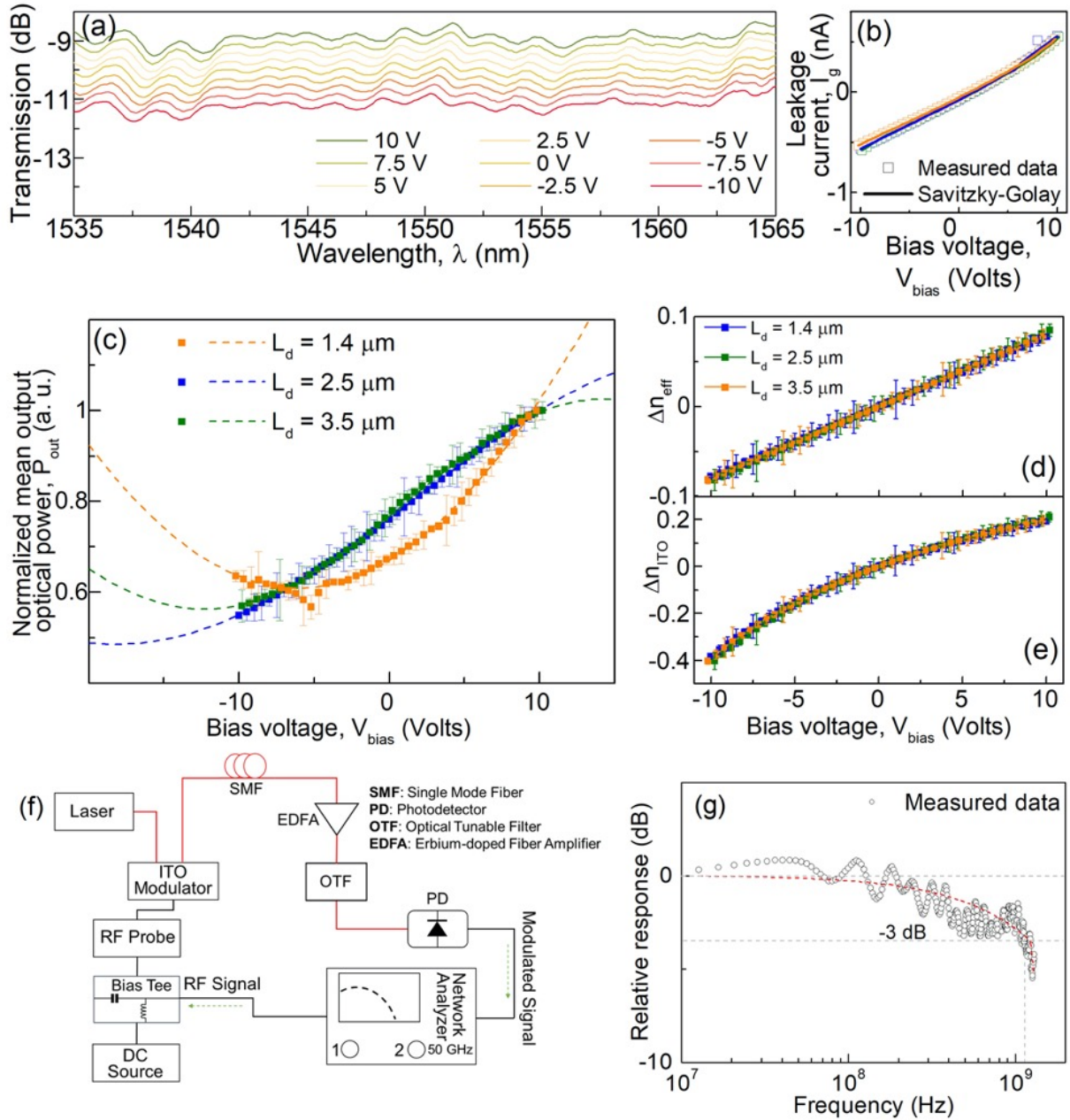

**Figure 2.** (a) Optical transmission exhibiting broadband performance of the modulator; (b) I-V measurements; (c) Optical output modulation under DC bias for different device scaling, dashed lines are represent $\cos^2(\text{arg})$ fit dictated by Mach-Zehnder operating principle; (d) Induced effective index change, $\Delta n_{eff}$; and (e) ITO material index change, $\Delta n_{ITO}$ from applied bias; (f) Experimental high-speed set-up; and (g) Measured small-signal response, $S_{21}$ of the modulator establishing a -3 dB bandwidth of 1.1 GHz.

In conclusion, here, we have demonstrated a spectrally broadband, wavelength-compact, GHz-fast ITO-based MZI modulator in silicon photonics showing a low $V_\pi L$ of 95 V·μm enabled by: a) efficient material modulation in ITO optimized for real-part index operation, b) a plasmonic hybrid mode enhancing the light-matter interaction, c) relatively low electrical resistance, and d) operating away from an optical resonance. This demonstration bears relevance, since ITO is a foundry-near material with a reduced barrier for co-integration. Future improvements on capacitance, electrostatics, and push-pull designs anticipate a ~10 GHz device with a few fJ/bit operation. Next-generation modulators, especially those with near-foundry potential, could be utilized not only in interconnect applications, but also for network-on-chip solutions [11] or more recently as nonlinear activation functions (threshold) in photonic artificial neural networks [12-15] for applications in machine learning. However, for the latter, the optical-electrical-optical conversion at every neuron using the modulator neurons could be improved with respect to signal delay by transitioning to all-optical neural networks [16].

**Acknowledgements**
Air Force Office of Scientific Research (FA9550-17-1-0071 and FA9550-17-1-0377).